\documentclass[preprint]{aastex}
\def\simgr{\,\hbox{\hbox{$ > $}\kern -0.8em \lower 1.0ex\hbox{$\sim$}}\,}
\def\simle{\,\hbox{\hbox{$ < $}\kern -0.8em \lower 1.0ex\hbox{$\sim$}}\,}

\shortauthors{PATTERSON ET AL.}
\shorttitle{CATACLYSMIC VARIABLE 1RXS J070407+262501}

\begin{document}

\title{Rapid Oscillations in Cataclysmic Variables.  XVII. 1RXS J070407+262501}

\author{Joseph Patterson\altaffilmark{1}, 
John R. Thorstensen\altaffilmark{2}, 
Holly A. Sheets\altaffilmark{2, 3},
Jonathan Kemp\altaffilmark{4,1},
Laura Vican\altaffilmark{1},
Helena Uthas\altaffilmark{5},
David Boyd\altaffilmark{6},
Michael Potter\altaffilmark{7},
Tom Krajci\altaffilmark{8},
Tut Campbell\altaffilmark{9}, 
George Roberts\altaffilmark{10}, 
Donn Starkey\altaffilmark{11}, 
and Bill Goff\altaffilmark{12}}

\altaffiltext{1}{Department of Astronomy, Columbia University, 550 West 120th Street, New York, NY 10027}
\altaffiltext{2}{Department of Physics and Astronomy, Dartmouth College, 6127 Wilder Laboratory, Hanover, NH 03755}
\altaffiltext{3}{Now at Department of Astronomy, University of Maryland
College Park, MD 20742-2421}
\altaffiltext{4}{Joint Astronomy Centre, University Park, 660 North A'ohoku Place, 
Hilo, HI 96720; j.kemp@jach.hawaii.edu}
\altaffiltext{5}{University of Southampton, Department of Physics and Astronomy, 
Highfield, Southampton SO17 1BJ, UK}
\altaffiltext{6}{CBA Oxford, 5 Silver Lane, West Challow, Wantage, Oxon, OX12 9TX, UK;
drsboyd@dsl.pipex.com}
\altaffiltext{7}{CBA Baltimore, 3206 Overland Ave, Baltimore, MD, 21214;
mike@orionsound.com}
\altaffiltext{8}{CBA New Mexico, PO Box 1351, Cloudcroft, NM, 88317;
tom\_krajci@tularosa.net}
\altaffiltext{9}{CBA Arkansas, 7021 Whispering Pine Road, Harrison, AR, 72601;
jmontecamp@yahoo.com}
\altaffiltext{10}{CBA Tennessee, 2007 Cedarmont Drive, Franklin, TN, 37067;
georgeroberts@comcast.net}
\altaffiltext{11}{CBA Indiana, 2507 County Road 60, Auburn, IN, 46706; donn@starkey.ws}
\altaffiltext{12}{CBA Sutter Creek, 13508 Monitor Lane, Sutter Creeek, CA, 95685;
b-goff@sbcglobal.net}

\begin{abstract}
We present a study of the recently discovered intermediate
polar 1RXS J070407+262501, distinctive for its large-amplitude
pulsed signal at $P = 480$ s.  Radial velocities indicate an
orbital period of 0.1821(2) d, and the light curves suggest
0.18208(6) d.  Time-series photometry shows a precise spin
period of 480.6700(4) s, decreasing at a rate of 0.096(9) ms yr$^{-1}$,
i.e. on a time scale $P/\dot P=2.5 \times 10^6$ yr.  The light curves
also appear to show a mysterious signal at $P = 0.263$ d, which
could possibly signify the presence of a ``superhump" in this
magnetic cataclysmic variable.
\end{abstract}

\keywords{keywords: stars}

\section{Introduction}

   IRXS J070407+262501 (hereafter RX0704) is a weak hard X-ray source
coinciding with a 16th magnitude star (USNO 1125.04825852).  
\citet[hereafter G05]{gaensicke05}
associated this star with the X-ray source,
and described its properties: a high-excitation cataclysmic variable 
(CV) with an orbital period
near 4 hours, and a strong optical pulse with a fundamental period of
480 s (and most of the power at 240 s, the first harmonic).  This suggests
membership among the DQ Hercuilis stars, or ``intermediate polars" as they
are also called \citep[for reviews and rosters, see][]{patterson94,hellier96, mukai09}.
This has now been confirmed by detection of X-rays pulsed
with the same period \citep{anzolin08}, which is interpreted as the
spin period of an accreting, magnetic white dwarf.

   The very high pulse amplitude caught our eye; we had been looking
for a star with a pulsed signal sufficiently strong that small-telescope
observers could track nearly every pulse -- and thereby accumulate a long
and detailed observational record of the periodicity.  This was partially
successful, although the star is a little faint, and the pulse a little
fast, to be tracked in fine detail by small telescopes.  Still, we managed
to learn the periods precisely, and to track long-term changes in pulse
period.  We report here on the results of that 2006-2010 campaign.

\section{Spectroscopy}

   We obtained spectra, using the 1.3 m McGraw-Hill telescope at MDM
Observatory, the Mark III spectrograph, a 600 line mm$^{-1}$ grism, and a SITe
1024 $\times$ 1024 CCD detector.  The spectra covered 4650-6980\ \AA\ with 2.3
\AA\ pixel$^{-1}$ and a slightly undersampled FWHM resolution of 4 
\AA\ over most of
the range.  The wavelength scale was established by frequent observation
of arc lamps.  The 2 arc-second slit produced light losses, so the flux
measurements are only approximate.

   We obtained a pair of 15 minute exposures on 2006 January 10, and much
more extensive observations during January 18-22.  Our observations span
a range of 9.4 hours in hour angle, which suppressed daily cycle count
ambiguities in the period determination.  Table 1 contains the log of
observations.

   Figure 1 shows the mean spectrum.  The synthetic magnitude, computed
using the passband from \citet{bessell90} is $V=17.3$.  Strengths of Balmer,
He I, and He II emission lines are given in Table 2.  Their equivalent
widths are typical of novalike variables, and especially those with
strong X-ray fluxes.  The spectrum is fairly similar to that of G05,
although our continuum flux level is nominally $\sim 1$ magnitude fainter.
We do not detect any absorption features from a secondary star.

   We measured radial velocities of the H$\alpha$ emission by convolving
its profile with the derivative of a Gaussian, optimized for a line
profile of 14 \AA\ FWHM.  \citet{schneider80} describe the
line-measurement algorithm.  The next strongest line, H$\beta$, also yielded
some useable velocities, but with lower signal-to-noise.  To search for
periods, we analyzed the H$\alpha$ velocities with the ``residual-gram"
method described by \citet{thorstensen96}.  This yielded an
unambiguous cycle count and a period of $262.2 \pm 0.4$ min.  This is
consistent with the less precise period suggested by G05.  The best
fit sinusoid of the form $v(t) = \gamma + K \sin[2\pi (t - T_0) / P]$ 
has 
\begin{eqnarray}
T_0 &= \hbox{HJD } 2,453,755.8430(29) \nonumber \\
P &= 0.1821(2) \hbox{ d} \\                               
K &= 191(12) \ \rm{km\ s^{-1}} \nonumber \\ 
\gamma &= 65(9) \ \rm{km\ s^{-1}}. \nonumber 
\end{eqnarray}
Fig.~2 shows this fit superposed on the folded H$\alpha$ velocities; note
that $T_0$ is the time of blue-to-red crossing.

\section{Photometry and Light Curves}

   We obtained time-series photometry on 43 nights, totaling 250 hours,
during 2006-2010.  About one-fifth consists of fast ($\sim$15 s resolution)
integrations with CCD cameras on the 1.3 m telescope of MDM Observatory.
The rest is slower (40-80 s) photometry with various 20-35 cm telescopes
in the Center for Backyard Astrophysics network \citep[CBA,][]{patterson98}.
About half the nights are quite long, $\sim$8 hr, and some are very
short ($\sim$1 hr, suitable only for a pulse timing).  The brightness
generally varied in the range $V$ =16.4-17.2.  Brightness changes up to 0.4
mag were noted on consecutive nights, but we could not detect any pattern
in these changes.  In this program of differential photometry, the quality
of the nights was mixed: some were photometric, but many were somewhat
cloudy, since we considered this star a good target for cloudy nights
(because we were primarily interested in the timing of its powerful and
fast pulse).
                               
    On every night and at all times, the star flashed a powerful 240 s
wave in the light curve, of $\sim$0.3 mag full amplitude (peak-to-trough).
This is shown in the upper frame of Fig.~3.  The lower frame shows the
power spectrum of one night's light curve.  This indicates a dominant
signal at 359.5 c d$^{-1}$ (towering off-scale at a power of 1600), an
obvious signal at lower frequency, and many harmonics of the latter.
This reveals the signal's underlying nature: a 480 s periodicity with a
double-hump waveform and other departures from a pure sinusoid.  All
of this agrees with the study of G05.

    On six occasions, spread over 40 nights, 
we obtained $\sim$8 hr coverage at high time resolution
with the 1.3 m telescope.  On each night, we folded the data on the
precise 480 s period; the resultant waveforms were identical each time.
Fig.~4 shows the waveforms for the two best 
pairs of consecutive 
nights (clear throughout),
demonstrating how precisely the waveform repeats.  Here we have defined
zero as the phase of the broader, deeper minimum -- which always precedes
the sharper maximum.  (This is different from the more practical phase
convention used below.)  Although only a few nights permitted study of
these subtle odd-even effects from the synchronous summations, we could
carry out the study for six nights by examining the harmonics.  Harmonics
carry information about the waveform, and we found that {\it the phasing of
the high harmonics was repeatable from night to night}.  This indicates
a repeatable waveform, at least on those six nights.

\section{Detailed Period Structure and Orbital Effects}

     The 2006-7 campaign primarily relied on MDM time series, while in 
2009-10 we had mainly CBA coverage.  We analyzed these two campaigns
separately.

\subsection{The 2009-10 Campaign}

     In 2009-10 we had only small telescopes available, but obtained 8
consecutive long nights of coverage from several terrestrial longitudes.
This overcame all problems of aliasing.  The upper frame of Figure 5
shows the low-frequency power spectrum, with significant signals flagged
with their frequency in cycles d$^{-1}$ ($\pm 0.015$).  The 5.479 c d$^{-1}$ signal is
apparently the orbital frequency, and the 3.805 c d$^{-1}$ signal is some
unknown longer-period wave (discussed in Sec.~7 below).  Inset is the
power-spectrum window, which demonstrates that there are no ambiguities
from aliasing.  The middle frame shows the waveforms of these two signals,
with an arbitrary zero-point (see caption).  The bottom frame shows other
segments of the 8-night power spectrum.  Note that the immediate vicinities
of the two strong signals at $\omega_{\rm spin}$ and $2\omega_{\rm spin}$ 
show the same picket-fence
pattern as the spectral window in the upper frame.  This indicates no
significant aliasing, and no systematic amplitude or frequency modulation.
There is a small but significant peak red-shifted by $5.477 \pm 0.015$ c d$^{-1}$
from $\omega_{\rm spin}$.  This is apparently the $\omega_{\rm spin} - 
\omega_{\rm orb}$ feature, a common
syndrome in DQ Her stars, probably arising from the reprocessing of pulsed
flux in structures fixed in the orbital frame (analogous to the
sidereal/synodic lunar month).

\subsection{The 2006-7 Campaign}

     In 2006-7 most of the coverage was with the MDM telescope, and
consisted of two 9-night clusters with a total span of 48 nights.  Each
cluster was somewhat sparse due to intervening bad weather, and there
were no data from distant longitudes.  The result was data of high signal-
to-noise and good definition of the periodic signals -- but fairly poor
rejection of aliases.  This nicely complemented the strengths of the
2009-10 campaign.                                     

     The light curves and power spectra were very similar to those of
2009-10.  In addition to the obvious powerful signals, a signal was
sometimes manifest near 5-6 c d$^{-1}$, and the lower orbital sideband of the
spin frequency near 174.25 c d$^{-1}$.  The latter provided some additional
and accurate measures of $\omega_{\rm orb}$
(assuming that the observed sideband of the spin
frequency is always separated by exactly $\omega_{\rm orb}$).  
These are listed in
Table 3.  The overall result, averaging over these
estimates, is $\omega_{\rm orb} = 5.492(2)$ c d$^{-1}$, 
or $P_{\rm orb} =
0.18208(6)$ d.

     This orbital period, mainly based on a 48-night baseline, is not
quite accurate enough to establish a unique cycle count during the 3-year
gap between campaigns.  However, one consecutive-night pair in 2008-9
yielded a candidate detection of the orbital signal.  With a likely
orbital maximum light at HJD 2454850.619, only one alias is permitted,
and the (candidate) orbital ephemeris becomes
$$\hbox{Orbital maximum} = 2454061.966(6) + 0.182052(3) E.$$

     The upper frame of Fig.~6 shows a 9-hour light curve, from which
the fast periodic wave (including harmonics) has been removed
\footnote{ 
Without this removal, slow variations are difficult to see -- masked by
the enormous strength of the pulsed signal.}.
A possible 5-6 hour wave is present, although the 2006-7 sampling was too
sparse to permit a reliable parsing into distinct periodic components at
low frequency.

    Superior signal-to-noise and time resolution in 2006-7 permitted
closer study of the fast signals.  Two consecutive nights with 8-hour
light curves gave the best determination.  The lower frame of Figure 6
shows the power spectrum after removal of the 480 s signal and all its
harmonics.  Significant signals are marked with their frequency in
cycles d$^{-1}$ ($\pm 0.08$), and each is a harmonic of $174.27 \pm 0.06$ 
c d$^{-1}$.  The
inset frame shows the mean light curve at this frequency, which is
$\omega_{\rm spin} - \omega_{\rm orb}$.

\section{Pulse Ephemeris and Long-Term Period Change}

    Our campaign also had coverage distributed over each observing season,
and hence is well suited for the study of long-term timing effects, since
an exact cycle count can be associated with each pulse.  Table 4 contains
the period derived for each observing season, along with the estimate of
G05 for the earlier 2004-5 season.  Basically, each season is consistent
with $P=480.6700(2)$ s, except for the first season, which is puzzling (see
also the $O-C$ analysis below).
     
    This period is easily accurate enough to count cycles from one year
to the next, and in Table 5 we list the times of maximum light derived
for each night.  Since the vast majority of the power occurs at the
240 s first harmonic, we fit each light curve with that period, and
extract the time of maximum light.  (Although we have cited evidence
above that the odd-even asymmetry is persistent, it is only provable on
the best nights, so a safer procedure is to use the powerful first
harmonic.)

    Figure 7 shows an $O-C$ diagram of these timings relative to a test
period of 240.335031 s.  The 48 points are fit by a simple parabola with
an RMS residual of 0.03 cycles (= 8 s), within measurement error.  The
best fit corresponds to 
\begin{equation}
\hbox{HJD pulse maximum} = 2454779.89661(3) + 0.0027816553(4)  E - 4.2(3) \times 10^{-15} E^2.
\end{equation}
Also shown, near $E = -5 \times 10^5$, are the three timings published by G05,
with cycle counts $E$ assigned to most closely match our ephemeris.  
These are hard to understand in view of the orderly behavior during 2006-2010.
Because of the very large scatter and odd period (see Table 4), we exclude
these points in the fit.  The parabolic term corresponds to $dP/dt = -0.095$
ms yr$^{-1}$, or a period decrease on a timescale 
$P / \dot P = 2.5 \times 10^6$ yr.

It is evident -- and fascinating -- that the reprocessed signal is 
dominated by the fundamental, while the spin
signal is obviously dominated by the first harmonic.
The spin signal likely
originates from gas falling radially onto the white dwarf, and the
strong harmonic presumably signifies a two-pole accretor, with both
poles in plain sight.  But in this noneclipsing star, structures
fixed in the orbital frame may see the two poles asymmetrically
(from our viewing angle, which will always favor one side of the
secondary, hot spot, eccretion disk, etc.)  That may well be the
basic explanation.  Fast spectroscopy of the emission lines may
provide an absolute phasing of the accretion funnel's location
with the photometric pulse, as done for AO Psc by \citet{hellier91}. 

\section{Comparison with X-rays}

    With a long-term spin ephemeris, we can compare the optical and X-ray
pulsed signals.  The X-ray waveform is complex, but distinguished by a
fairly sharp dip in soft X-rays, presumably arising from photoelectric
absorption in the accretion column.  \citet{anzolin08} report two
epochs for the absorption dip in 2006-7, which are separated, according
to our ephemeris, by $30635.01 \pm 0.05$  
cycles (of the 480 s period).  So the dip
location appears to be stable in phase.  During 2006-7, the ephemeris for
the broad minimum shown in Figure 5 is
$$\hbox{Broad minimum} = \hbox{HJD } 2454108.64254 + 0.005563318 E,$$
which implies that both X-ray absorption dips (each being an average
over $\sim$50 spin cycles) occurred at spin phase 0.88 on the convention of
Fig.~4.  We could probably have concocted some explanation for simple
fractions like 0.50, 0.00, and even 0.75.  But 0.88 is just too
challenging.  As usual with DQ Her stars, the comparison of optical and
X-ray phases -- though precise -- is mysterious.

\section{The 6.3 Hour Signal}

    A puzzling feature of the low-frequency power spectrum in Figure 5
is the 3.8 c d$^{-1}$ (= 6.3 hr) signal.  Its amplitude is similar to that of
the (obviously real) orbital signal, and the inset power-spectrum window
shows that it is not aliased to the orbital signal.  Nor does it occur
at any suspicious frequency -- e.g. 1/2/3 c d$^{-1}$, which can be artifacts
of differential extinction.  Thus it is likely to be a real signal which
maintained some coherence over the 8-night interval.

    Cataclysmic variables are well-equipped to produce signals at 
$\omega_{\rm orb}$,
due to eclipses, presentation effects of the orbiting secondary, an emitting
hot spot, and an absorbing hot spot.  But distinct signals at a nearby
frequency are difficult to understand.  The only such phenomenon which is
(moderately) understood is the ``superhump" - which probably arises from
a periodic modulation of accretion-disk light induced by the secondary's
perturbation at the disk's 3:1 resonance 
\citep[e.g.,][]{whitehurst91, smith07}.
About a hundred stars are known
to show superhumps, and about twenty papers have explored the theory
underlying the phenomenon.  But no confirmed
superhumpers are known to be magnetic, and no theory has allowed for that
possibility.  Yet in the case of RX0704, the rapid X-ray/optical pulse
certifies strong magnetism, {\it and} disruption of the disk by that magnetism
(in order to provide a radial plunge to the white-dwarf surface).  Is it
possible that such a star can also show superhumps?

    Yes, perhaps.  \citet{retter01,retter03} reported a 6.3 hr
photometric signal in the 5.5 hr binary TV Columbae, which is also a
DQ Her star.  Retter et al.~considered this to be a superhump, although
this has not been widely accepted, because other observations of comparable
sensitivity failed to show the signal (and perhaps because there was no
precedent among magnetic CVs).  However, it's common among confirmed
superhumpers that the signal can be somewhat transient -- present and powerful
in one long observing campaign, and missing in another.  

But the case in RX0704 is still doubtful.  This is just one weak
detection -- and, for that matter, a detection over just one week
(8 nights).  Even though it is not a simple alias, a weak signal at
low frequency can potentially be produced by amplitude modulations of the
orbital signal, or by more complex artifacts of the data-taking -- e.g.,
arising from an interaction of differential extinction with the ``hand-off"
between different telescopes in this multi-longitude campaign.  Such
worries can be mitigated by a much longer observing campaign, or
by the star's decision to flash a signal of greater amplitude.
                
\section{Summary}

1.  From radial velocities and photometry, we establish a precise binary
period of 0.18208(5) d.  This period is also manifest as a weak photometric
signal, and as the lower orbital sideband 
($\omega_{\rm spin} - \omega_{\rm orb}$) of the main
pulse frequency, probably indicating a component reprocessed in structures
fixed in the orbital frame.

2. The white dwarf spins with P = 480.6700 s, with most power at the first
harmonic, very likely signifying a two-pole accretor.  Oddly, the signal at the lower
orbital sideband frequency (``reprocessed component") appears to be dominated
by the fundamental.

3. The X-ray dip
occurs 0.12 cycles before the broader optical minimum. 

4. The spin period decreases on a timescale of $P/\dot P = 2.5 \times 10^6$ yr.

5. There may be a low-frequency photometric signal at P = 6.3 hours.
Future observation should seek to clarify whether this is a common feature
of the star.  If confirmed, this periodic wave might be a long-period 
manifestation of a ``superhump", with the additional novelty that it occurs 
in a magnetic cataclysmic variable.  Further theoretical exploration of 
this possibility is very desirable.

\acknowledgments
    We gratefully acknowledge support from the National Science
Foundation (through AST-0908363 at Columbia, and grants AST-0307413 
and AST-0708810 at Dartmouth), NASA grant GO11621.03a, and the Mount
Cuba Observatory Foundation.

Facilities: \facility{McGraw-Hill}

\clearpage

\begin{deluxetable}{lrrrrr}
\tablewidth{0pt}
\tablecolumns{6}
\tablecaption{Radial Velocities and Hour Angles}
\tablehead{
\colhead{Time \tablenotemark{a}} &
\colhead{$V$ (H$\alpha$)} &
\colhead{$\sigma_V$} &
\colhead{$V$ (H$\beta$)} & 
\colhead{$\sigma_V$} &
\colhead{HA} \tablenotemark{b}\\
&
\colhead{(km s$^{-1}$)} &
\colhead{(km s$^{-1}$)} &
\colhead{(km s$^{-1}$)} &
\colhead{(km s$^{-1}$)} & 
\colhead{(hh:mm)} \\
}
\startdata
53745.9704  & $  -38$ & $  40$ & $ -108$ & $  60$  & +3:58  \\  
53745.9875  & $  -97$ & $  42$ & $  -45$ & $ 102$  & +4:22  \\  
53753.8997  & $  180$ & $  21$ & $   69$ & $  72$  & +2:47  \\  
53753.9082  & $  199$ & $  23$ &  \nodata & \nodata  & +3:00  \\  
53753.9167  & $  157$ & $  22$ & $  208$ & $  51$  & +3:12  \\  
53753.9253  & $   84$ & $  31$ & $  143$ & $  69$  & +3:24  \\  
53753.9338  & $   74$ & $  19$ & $  185$ & $  41$  & +3:36  \\  
53753.9443  & $   12$ & $  32$ & $   -1$ & $  39$  & +3:52  \\  
53753.9528  & $  -59$ & $  30$ & $    9$ & $  86$  & +4:04  \\  
53753.9613  & $  -30$ & $  44$ & $  102$ & $  43$  & +4:16  \\  
53753.9698  & $  -91$ & $  25$ & $   75$ & $ 193$  & +4:28  \\  
53753.9784  & $  -64$ & $  36$ & $  -36$ & $  87$  & +4:41  \\  
53754.7713  & $  175$ & $  19$ & $  215$ & $ 144$  & $-$0:14  \\  
53754.7819  & $  254$ & $  25$ & $  227$ & $  53$  & +0:01  \\  
53754.7904  & $  291$ & $  21$ &  \nodata & \nodata  & +0:13  \\  
53754.7990  & $  254$ & $  25$ &  \nodata & \nodata  & +0:26  \\  
53754.8075  & $  225$ & $  35$ & $  289$ & $ 100$  & +0:38  \\  
53754.8160  & $  126$ & $  51$ & $  377$ & $ 137$  & +0:50  \\  
53755.9046  & $   56$ & $  64$ & $  135$ & $ 106$  & +3:02  \\  
53755.9131  & $  104$ & $  66$ & $  182$ & $ 109$  & +3:15  \\  
53755.9216  & $   -8$ & $  57$ & $   59$ & $ 118$  & +3:27  \\  
53755.9408  & $   31$ & $  48$ & $  -10$ & $ 210$  & +3:55  \\  
53755.9493  & $  -62$ & $  64$ & $   41$ & $ 124$  & +4:07  \\  
53755.9578  & $ -204$ & $  82$ & $  -20$ & $  60$  & +4:19  \\  
53755.9776  & $ -250$ & $  47$ & $  -48$ & $ 122$  & +4:48  \\  
53755.9867  & $  -96$ & $  99$ & $   36$ & $  68$  & +5:01  \\  
53756.9172  & $  -54$ & $  29$ & $   83$ & $  49$  & +3:24  \\  
53756.9257  & $   24$ & $  22$ & $  123$ & $  62$  & +3:37  \\  
53756.9342  & $   41$ & $  26$ & $   56$ & $  49$  & +3:49  \\  
53756.9427  & $   43$ & $  24$ & $   67$ & $  74$  & +4:01  \\  
53756.9532  & $  139$ & $  20$ & $  197$ & $  46$  & +4:16  \\  
53756.9617  & $  217$ & $  15$ & $  311$ & $  51$  & +4:29  \\  
53756.9702  & $  270$ & $  17$ & $  325$ & $  73$  & +4:41  \\  
53756.9787  & $  277$ & $  24$ & $  316$ & $  41$  & +4:53  \\  
53756.9872  & $  224$ & $  22$ & $  248$ & $  54$  & +5:06  \\  
53757.5922  & $ -111$ & $  30$ & $  -32$ & $  50$  & $-$4:21  \\  
53757.6007  & $  -23$ & $  42$ & $ -143$ & $ 132$  & $-$4:09  \\  
53757.6093  & $ -210$ & $  36$ & $ -256$ & $  88$  & $-$3:56  \\  
53757.6178  & $  -55$ & $  33$ & $   37$ & $ 129$  & $-$3:44  \\  
53757.6263  & $ -130$ & $  28$ & $    7$ & $  79$  & $-$3:32  \\  
53757.6433  & $  -61$ & $  24$ & $   48$ & $ 204$  & $-$3:07  \\  
53757.6518  & $  -26$ & $  29$ & $  -39$ & $  54$  & $-$2:55  \\  
53757.6603  & $   29$ & $  26$ & $  140$ & $  69$  & $-$2:43  \\  
53757.6688  & $   58$ & $  24$ & $   42$ & $  57$  & $-$2:30  \\  
53757.6773  & $  192$ & $  28$ & $  129$ & $  54$  & $-$2:18  \\  
53757.6887  & $  242$ & $  22$ & $  208$ & $  78$  & $-$2:02  \\  
53757.6972  & $  263$ & $  19$ & $  235$ & $  88$  & $-$1:49  \\  
53757.7058  & $  292$ & $  16$ & $  353$ & $  34$  & $-$1:37  \\  
\enddata
\tablenotetext{a}{Heliocentric Julian date of mid-integration,
minus 2400000. Integrations were 720 and 900 s, and the time
system is UTC.}
\tablenotetext{b}{Hour angle at mid-integration.}
\label{t:velocities}
\end{deluxetable}

\clearpage

\begin{deluxetable}{lrcc}
\tablewidth{0pt}
\tablecolumns{4}
\tablecaption{Spectral Features in Quiescence}
\tablehead{
&
\colhead{E.W.\tablenotemark{a}} &
\colhead{Flux}  &
\colhead{FWHM \tablenotemark{b}} \\
\colhead{Feature} &
\colhead{(\AA )} &
\colhead{(10$^{-16}$ erg cm$^{-2}$ s$^{1}$)} &
\colhead{(\AA)} \\
}
\startdata
 HeII $\lambda 4686$ & $ 10$ & $ 55$ & 17 \\ 
            H$\beta$ & $ 12$ & $ 61$ & 18 \\ 
  HeI $\lambda 4921$ & $  1$ & $  6$ & 17 \\ 
  HeI $\lambda 5015$ & $  1$ & $  5$ & 20 \\ 
 HeII $\lambda 5411$ & $  3$ & $ 14$ & 26 \\ 
  HeI $\lambda 5876$ & $  3$ & $ 13$ & 18 \\ 
           H$\alpha$ & $ 26$ & $ 88$ & 20 \\ 
  HeI $\lambda 6678$ & $  3$ & $  9$ & 24 \\ 
\enddata
\tablenotetext{a}{Emission equivalent widths are counted as positive.}
\tablenotetext{b}{From Gaussian fits.}
\label{t:emissionlines}

\end{deluxetable}

\begin{deluxetable}{ll}
\tablewidth{0pt}
\tablecolumns{2}
\tablecaption{Accurate Estimates of the Orbital Frequency}
\tablehead{
\colhead{Source} &
\colhead{Frequency} \\
\colhead{} & 
\colhead{[c d$^{-1}$]} \\
}
\startdata
Radial velocities  &    5.492(7)  \\
2007a              &    5.495(14) \\
2007a (sideband)   &    5.495(13) \\
2007b              &    5.487(12) \\
2007b (sideband)   &    5.490(13) \\
2007a+b            &    5.494(4) \\
2007a+b (sideband) &    5.492(3) \\
2010               &    5.479(20) \\
2010 (sideband)    &    5.477(20) \\
ALL                &    5.492(2) \\
\enddata
\end{deluxetable}

\clearpage

\begin{deluxetable}{lll}
\tablewidth{0pt}
\tablecolumns{3}
\tablecaption{Yearly Estimates of the Pulse Period}
\tablehead{
\colhead{Season} &
\colhead{$P$ [s]} &
\colhead{Reference} \\
}
\startdata
2004-5  &  480.7080(44) &   G05 \\
2006-7  &  480.6712(10) &   this work \\
2007-8  &  480.6686(14) &   this work \\
2008-9  &  480.6695(6)  &   this work \\
2009-10 &  480.6700(3)  &   this work \\
\enddata 
\end{deluxetable}

\begin{deluxetable}{rrrrr}
\tablewidth{0pt}
\tablecolumns{5}
\tablecaption{Times of 240 s Maximum Light}
\tablehead{
\multicolumn{5}{c}{[Heliocentric Julian Days minus 2454000.]}
}
\startdata
  61.85645  &    63.81754 &    65.74801 &    69.82591 &    70.80234 \\
 100.68293  &   108.64128 &   109.75666 &   170.74176 &   179.75711 \\
 447.86435  &   474.95487 &   475.93117 &   779.89672 &   850.56729 \\
 851.56040  &   903.64976 &   904.62324 &   913.64979 &  1160.87467 \\
1164.70780  &  1166.66626 &  1176.68833 &  1177.67862 &  1182.66893 \\
1194.74976  &  1195.85697 &  1197.87922 &  1199.71220 &  1200.71092 \\
1201.30060  &  1201.78466 &  1202.74428 &  1203.30057 &  1205.58990 \\
1206.69426  &  1260.58046 &  1261.58466 &  1270.38303 &  1270.67511 \\
1271.64317  &  1305.61554 &  1312.62812 &  1470.85121 &  1471.84707 \\
1478.90405  &  1479.90545 &  1480.95974 &  1481.83037 \\
\enddata
\end{deluxetable}

\clearpage

\begin{figure}
\plotone{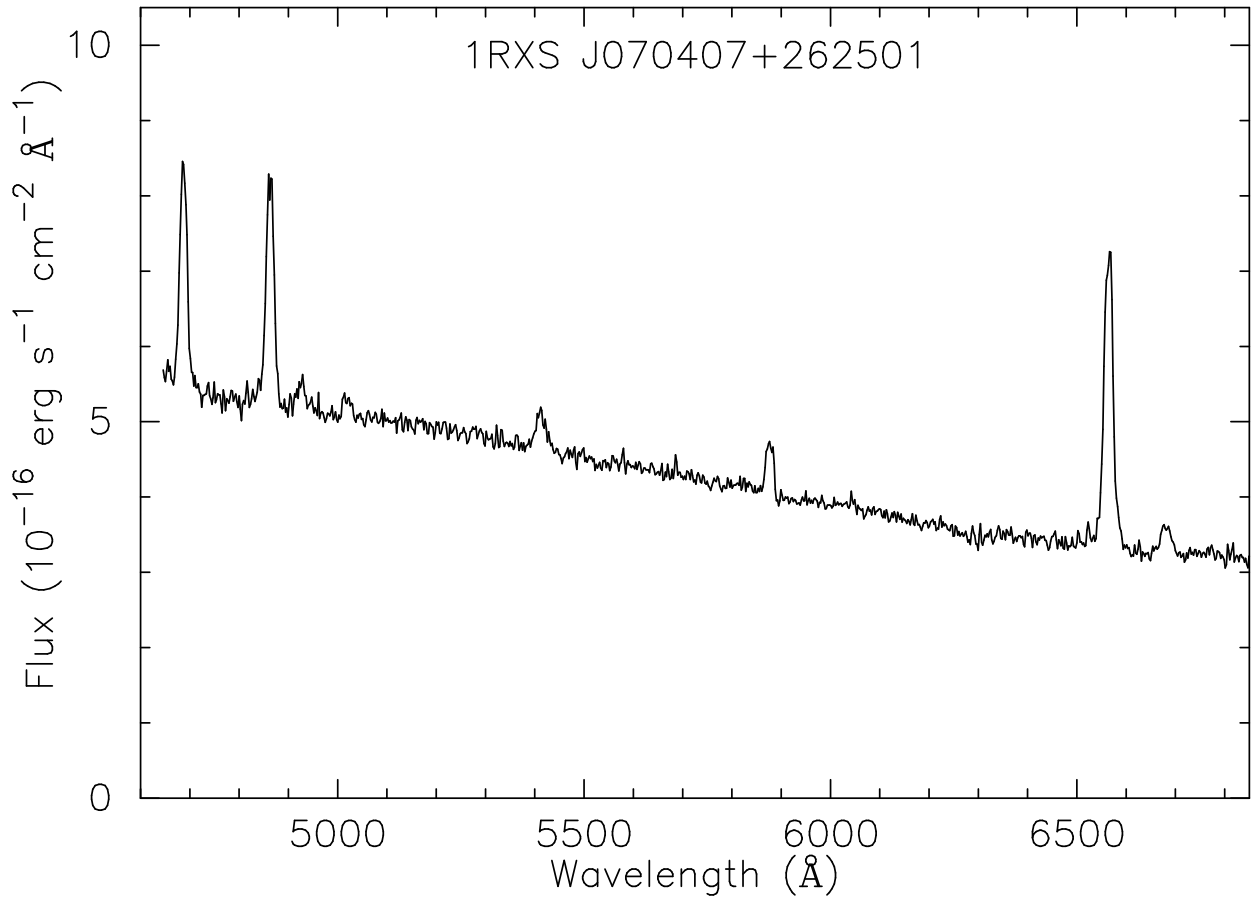}
\caption{Average flux-calibrated spectrum during 2006 January.  The
calibration of the vertical scale is uncertain by at least 20\%.}
\end{figure}

\clearpage

\begin{figure}
\plotone{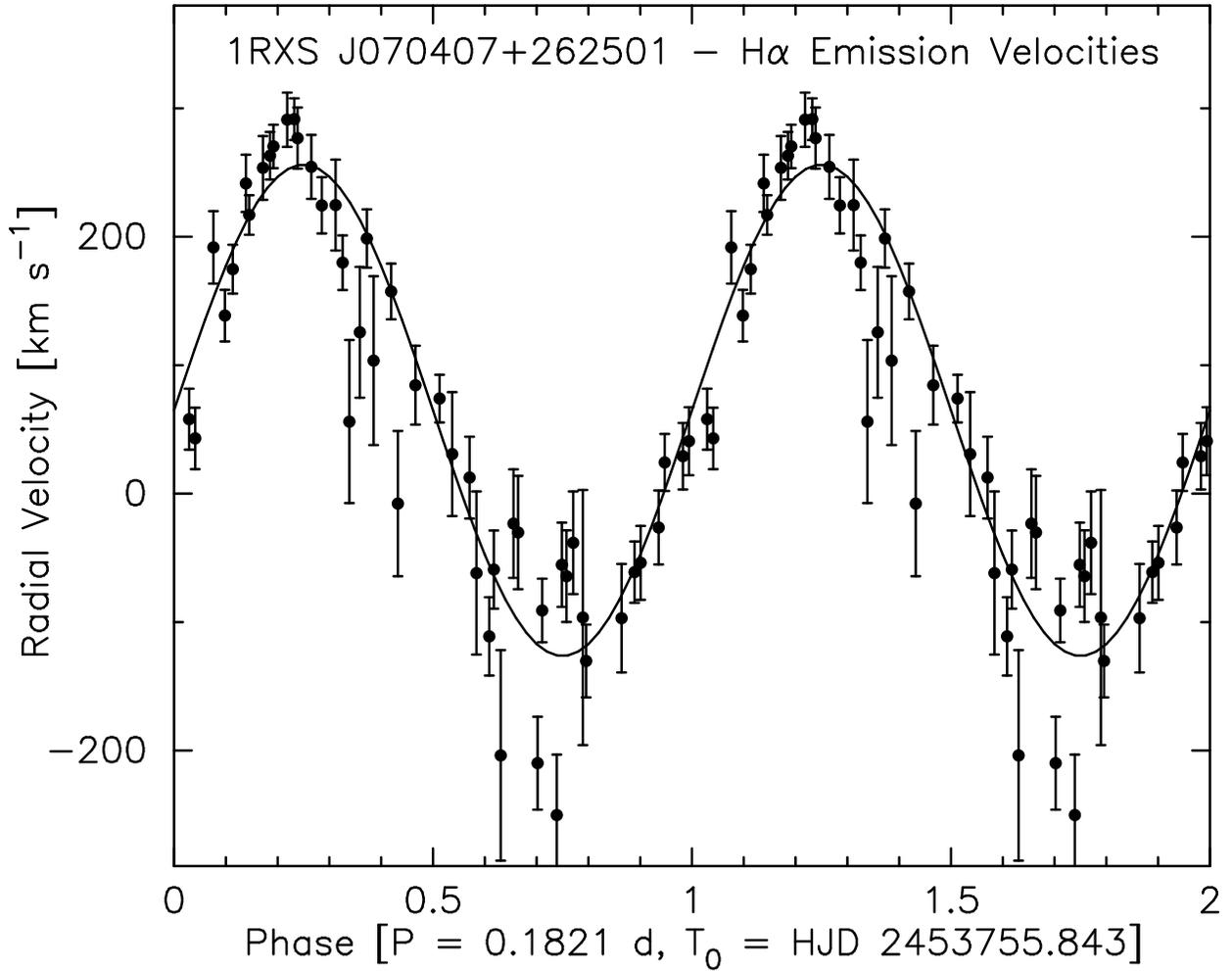}
\caption{
Radial velocities of the H$\alpha$ emission line, folded on the
spectroscopic period, and with the best-fit sinusoid superposed.  For
continuity, all data are repeated for a second cycle.  The errror bars
shown are 1 standard deviation, estimated from counting statistics.
}
\end{figure}

\clearpage

\begin{figure}
\plotone{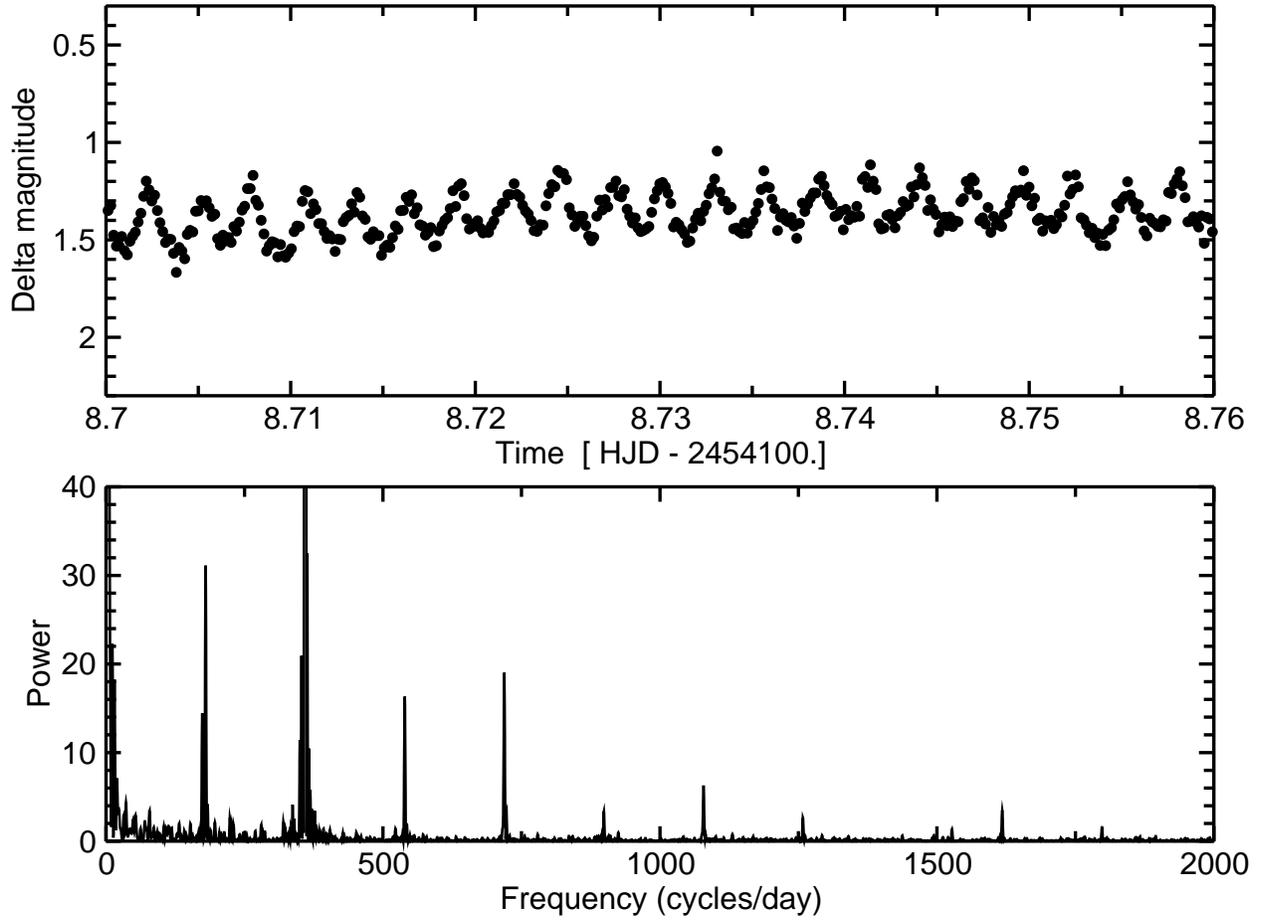}
\bigskip\bigskip
\caption{
{\it Upper frame:} A portion of one night's light curve, showing
the large 240 s pulses.  {\it Lower frame:} The power spectrum of the same
light curve.  Rising far off-scale (to a power of 1600) is the 240 s
signal; other features are the 480 s fundamental and many harmonics.
}
\end{figure}

\clearpage

\begin{figure}
\epsscale{0.95}
\plotone{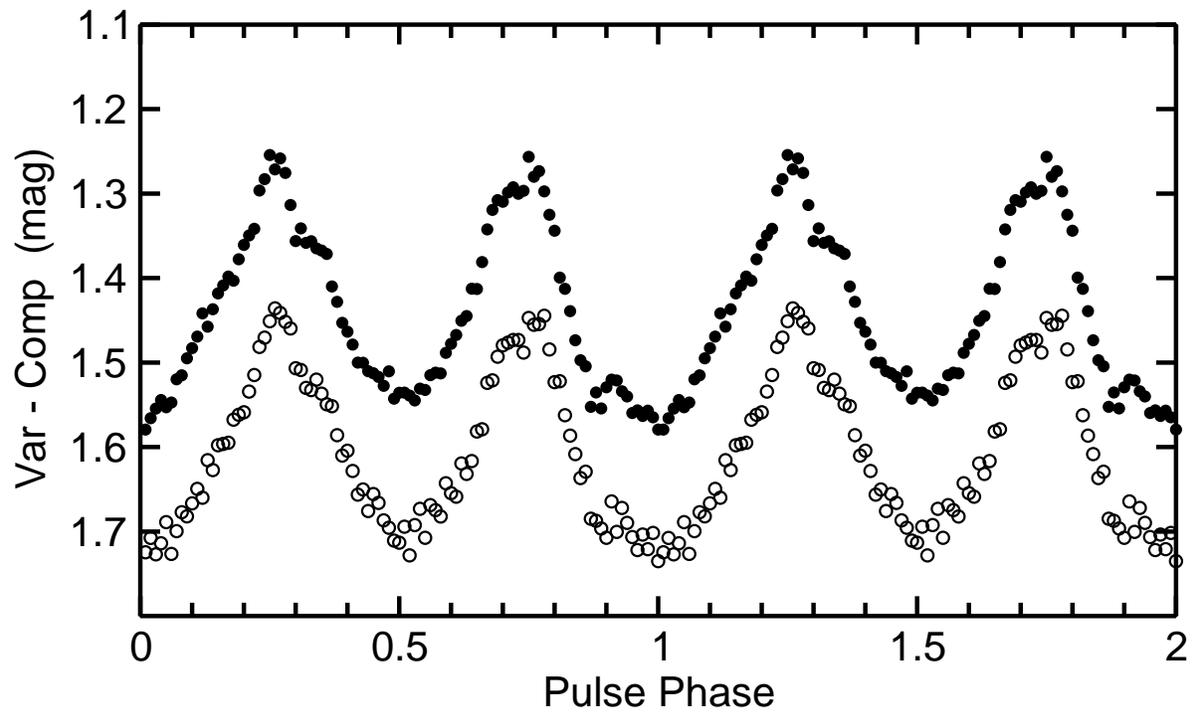}
\caption{
Mean light curve during two pairs of consecutive nights in
2006-7.  Zero phase is defined here as the phase of the broader, deeper
minimum in the 480 s cycle, and occurs at HJD 2454108.64254.
}
\end{figure}

\clearpage
\begin{figure}
\epsscale{0.77}
\plotone{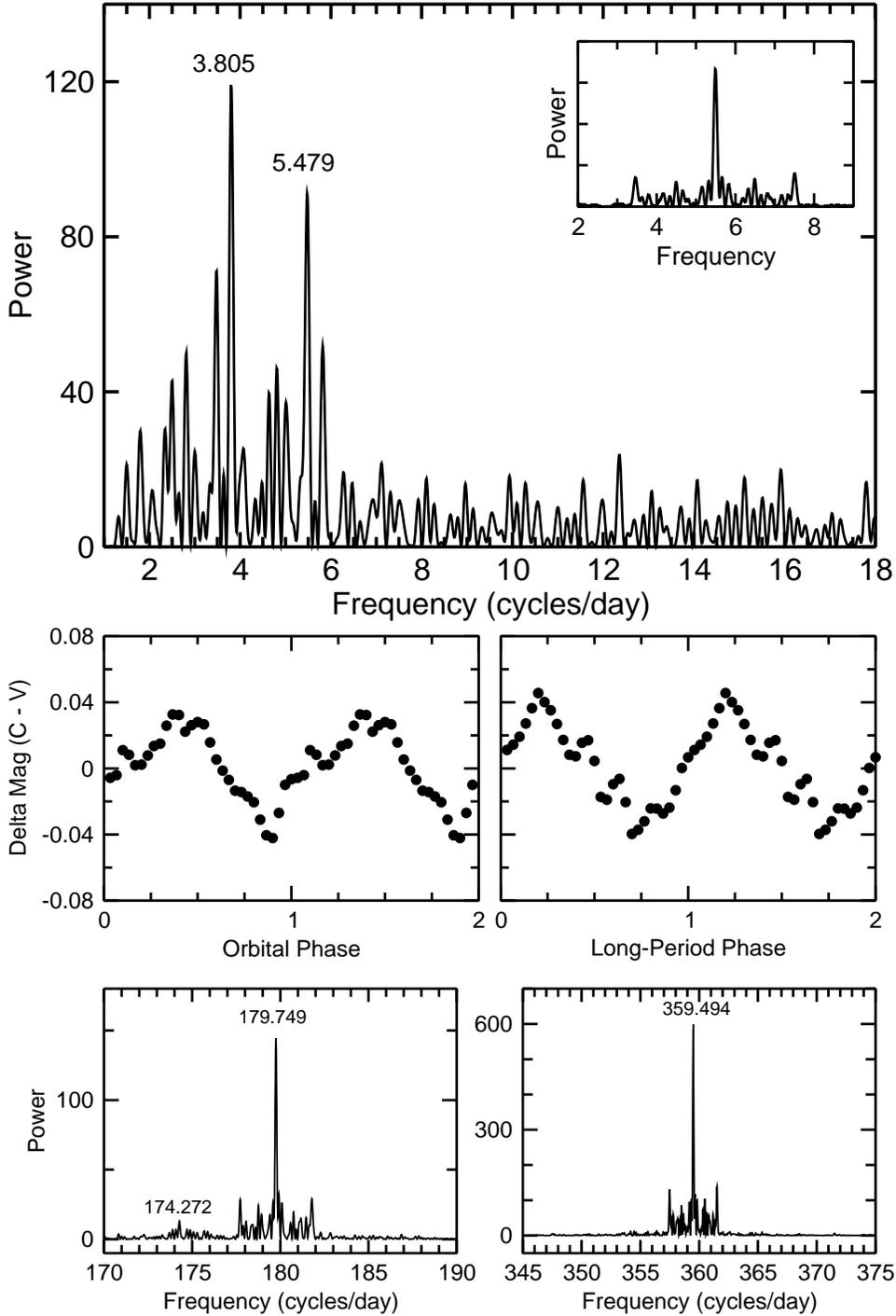}
\caption{
Eight-night power spectrum in 2009-10.  {\it Upper frame:} the
low-frequency region, with significant features flagged with their
frequencies in cycles d$^{-1}$ ($\pm 0.015$).  Inset is the power-spectrum window
of this campaign, which demonstrates that there are no problems with
aliasing - and, in particular, that the peculiar signal at 3.805 c d$^{-1}$ is
not an alias. {\it Middle frame:} Light curves folded at these periods: each
of semi-amplitude 0.04 mag.  {\it Lowest frame:} Region near the strong
signals, showing a significant peak at 174.27 c d$^{-1}$ 
($\omega_{\rm spin} - \omega_{\rm orb}$).
}
\end{figure}

\begin{figure}
\epsscale{0.82}
\plotone{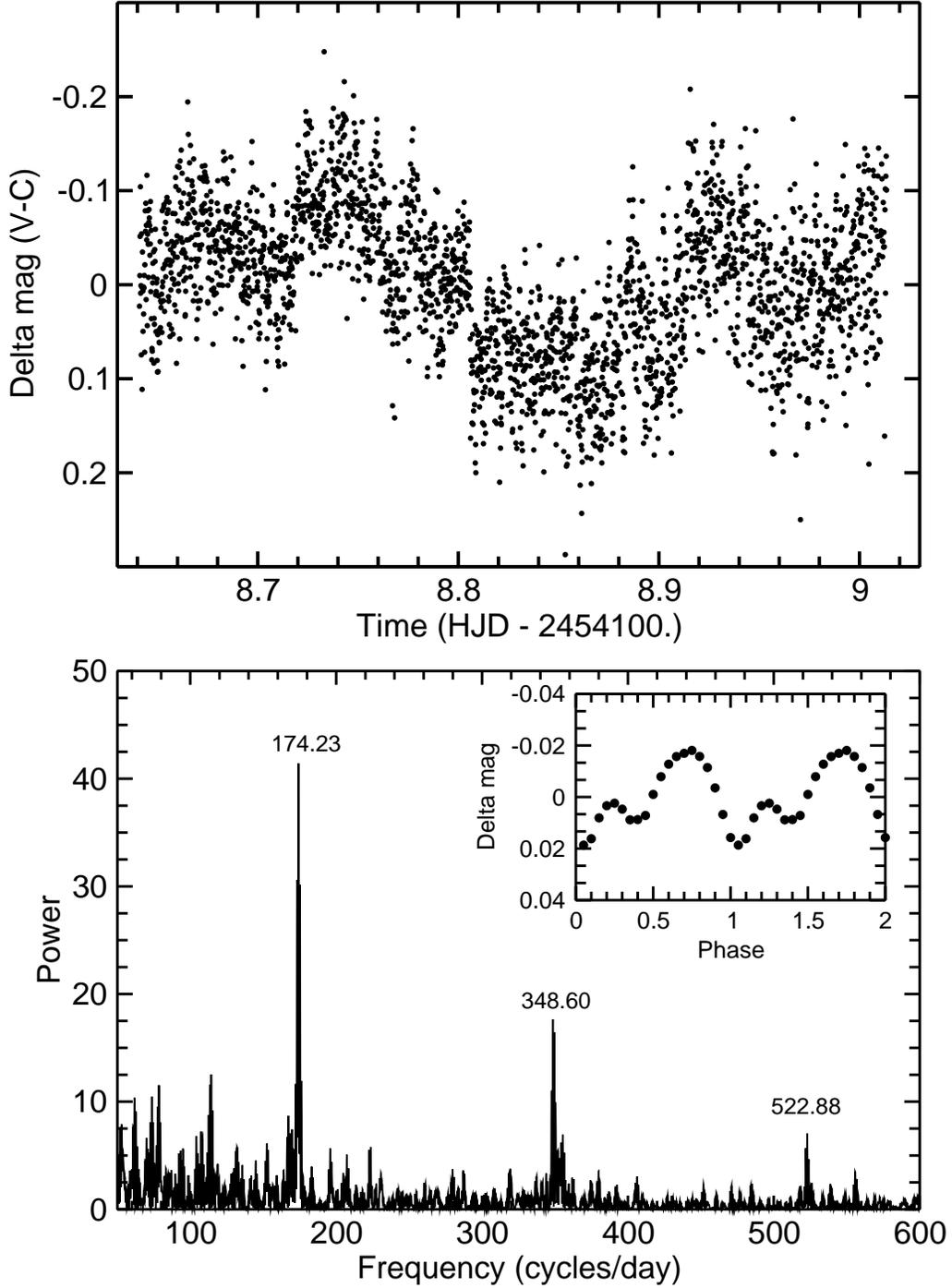}
\caption{
{\it Upper frame:} A 9-hour light curve after the rapid pulse is
removed.  {\it Lower frame:} The power spectrum of two consecutive
nights after that removal.  Each
significant peak is labelled with its frequency ($\pm 0.08$) in cycles d$^{-1}$.
They are integer multiples of $174.25 \pm 0.05$, which we identify as
($\omega_{\rm spin} - \omega_{\rm orb}$).  Inset is the mean waveform
of that sideband signal.  It is relatively weak, and is dominated
by the fundamental.
}
\end{figure}

\begin{figure}
\epsscale{1.0}
\plotone{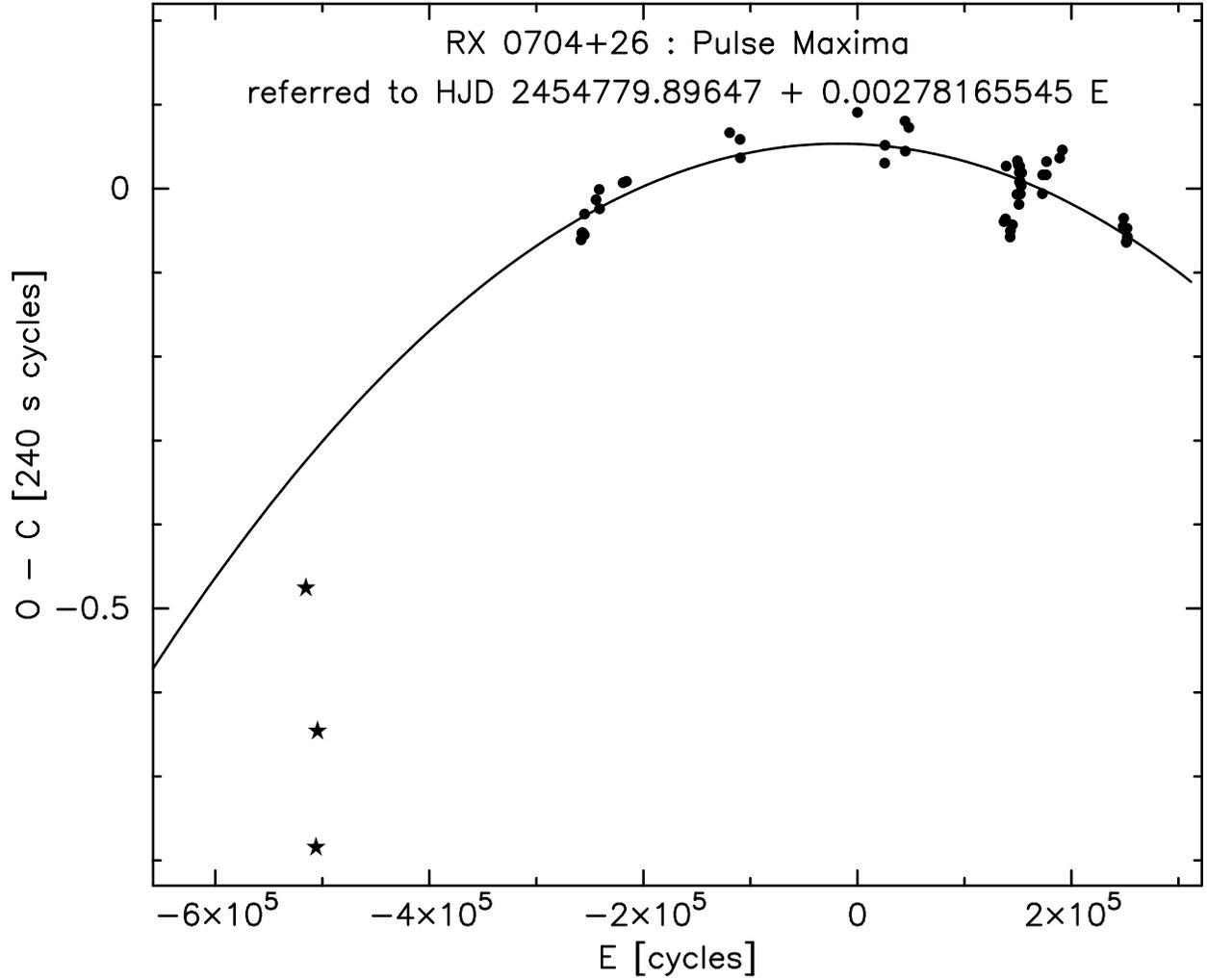}
\caption{
Figure 7. O-C diagram of the pulse maxima, relative to the test ephemeris
HJD 2454779.89647 + 0.00278165545 $E$.  The curve corresponds to the fitted parabola,
Eq. (2).  The three points near $E=-5 \times 10^5$ (stars) are those of G05; these are not
included in the fit.
}
\end{figure}

\end{document}